\documentstyle[twocolumn,aps]{revtex}
\tighten
\draft
\begin{document}
\title{Normal-State Hall Effect and the Insulating Resistivity \\
of High-$T_c$ Cuprates at Low Temperatures}
\author{Yoichi Ando,$^{1,2}$ G. S. Boebinger,$^{1}$ A. Passner,$^{1}$
N. L. Wang,$^{3}$ C. Geibel,$^{3}$ F. Steglich,$^{3}$\\
I.E. Trofimov,$^{4,}$\cite{byline} and F. F. Balakirev$^{4}$}
\address{$^{\rm 1}$ Bell Laboratories, Lucent Technologies, 700 Mountain
Avenue, Murray Hill, NJ 07974}
\address{$^{\rm 2}$ Central Research Institute of Electric Power Industry,
Komae, Tokyo 201, Japan}
\address{$^{\rm 3}$ Institut f\"ur Festk\"orperphysik, Technische
Hochschule Darmstadt, D-64289 Darmstadt, Germany}
\address{$^{\rm 4}$ Department of Physics and Astronomy, Rutgers University,
Picataway, NJ 08855}
\date{Received hall7.tex submitted}
\maketitle

\begin{abstract}
The normal-state Hall coefficient $R_H$ and the in-plane
resistivity $\rho_{ab}$ are measured in La-doped
${\rm Bi_{2}Sr_{2}CuO_{y}}$ ($T_c \simeq 13$ K) single crystals and
${\rm La_{2-x}Sr_{x}CuO_{4}}$ thin films by suppressing
superconductivity with 61-T pulsed magnetic fields.
In contrast to data above $T_c$, the $R_H$ below $\sim10$ K shows
little temperature dependence in all the samples measured, irrespective of
whether $\rho_{ab}$ exhibits insulating or metallic behavior.
Thus, whatever physical mechanism gives rise to insulating behavior in
the low-temperature normal state, it leaves the Hall conductivity
relatively unchanged.
\end{abstract}

        

Application of a pulsed high magnetic field to suppress
superconductivity in the high-$T_c$ cuprates has opened up the
possibility for measurements of normal-state transport at low temperatures.
This regime has been rather unexplored due to the extremely high $H_{c2}$ of
the cuprates.
Thus far the anisotropic normal-state resistivity has been measured in
${\rm La_{2-x}Sr_{x}CuO_{4}}$ (LSCO) \cite{logT-PRL,x-dep-PRL} and
La-doped ${\rm Bi_{2}Sr_{2}CuO_{y}}$ (Bi-2201) \cite{Bi2201-PRL}
down to subkelvin temperatures using 61-T pulsed magnetic fields.

One of the surprising findings in the low-temperature normal-state
resistivity of LSCO is an unusual $\log (1/T)$ divergence of both in-plane
($\rho_{ab}$) and $c$-axis resistivity ($\rho_c$) of underdoped samples.
Similar divergence of resistivity that is consistent with $\log (1/T)$
is also found in
disordered Bi-2201 \cite{Bi2201-PRL}, where the dynamic range
of the divergence is generally smaller than in underdoped LSCO, 
where the increase of $\rho_{ab}$ can be as large as a factor of three.
Although the origin of the unusual $\log (1/T)$ behavior is not yet clear,
there are several proposed explanations, some of which involve the
localization of the charge-carrying quasi-particles
\cite{Anderson,Alexandrov}, others of which involve the
suppression of the two-dimensional density of states near the Fermi
energy.  In this latter scenario, the low-temperature insulating behavior
in the cuprates might result
from conventional disorder-enhanced electron interactions \cite{Altshuler},
from the temperature-dependent impurity scattering time in the marginal
Fermi liquid \cite{Varma}, or from the 
existence of a pseudo-gap in the underdoped
high-$T_c$ cuprates \cite{Pines}.

In conventional physics of disordered metals in two-dimensions, both
weak localization and electron-electron interactions give rise to
identical $\log (1/T)$ corrections to the resistivity.  Measurement of
the Hall coefficient $R_H$ is useful in separating these two physical
mechanisms \cite{Altshuler}: weak localization does not affect $R_H$,
while interactions lead to corrections in $R_H$ which are
two times larger than the corrections to the resistivity.  The physics of
the cuprates is expected to be very different from that of conventional
disordered metals.  Indeed, three independent reports of logarithmic
behavior in underdoped cuprates provide three separate arguments against
interpretations involving conventional weak
localization \cite{logT-PRL,Preyer,Jing}.  Nevertheless, measurement
of the Hall effect down to low temperatures could help resolve which
of the proposed physical mechanisms governs low-temperature transport
in the cuprates.

In this paper, we present low-temperature measurements of $\rho_{ab}$
and the Hall coefficient, $R_H \equiv \rho_{\rm Hall} / B$, in Bi-2201 single
crystals and LSCO thin films using 61-T pulsed magnetic fields to suppress
superconductivity.  For Bi-2201, we measured several crystals with
nominally the same carrier concentration, for which the low-temperature
behavior of $\rho_{ab}$ is either metallic ($d \rho_{ab} /dT>0$) or
insulating ($d \rho_{ab} /dT<0$), depending upon the apparent amount of
naturally-occurring disorder in the different samples.  The LSCO thin
films have very different carrier concentrations, such that the
low-temperature $\rho_{ab}$ behavior also varies from metallic to
insulating between the samples.

The Bi-2201 single crystals are grown by the flux method in
${\rm Al_2O_3}$ crucibles \cite{Wang-xtal}.
To obtain nearly optimally doped crystals, La is doped onto the Sr site,
giving the composition of ${\rm Bi_{2}Sr_{2-x}La_{x}CuO_{y}}$ with
nominal $x$=0.05 and $T_c$ (midpoint) around 13 K. The actual La
concentration is estimated to be much higher, $x \sim$0.3, corresponding
to slightly overdoped samples \cite{Wang-dope}. The samples are annealed in
flowing oxygen for 1 hour at 400 C after six silver contacts are painted
on the edges of the platelet-shaped crystals.
Typical size of the crystals is
$1.7 \times 0.7 \times 0.02$ ${\rm  mm}^3$, small enough that the
data are not adversely affected by
eddy-current heating during the magnet pulses \cite{Bi2201-PRL}.

The LSCO films are c-axis-oriented, epitaxial films deposited on
${\rm LaSrAlO_4}$ substrates by pulsed laser deposition, as described in
Ref. \cite{Trofimov}.
One film is overdoped, with a nominal $x$=0.23 and $T_c$
(midpoint) of 19 K. The other is underdoped, with a nominal $x$=0.13. This
film was annealed in high-pressure oxygen \cite{Trofimov} yielding a
$T_c$ (midpoint) of 28 K.
The films were patterned by photolithography into a
conventional Hall bar with a center strip 0.2 mm wide. Silver contact
pads were evaporated onto the arms and gold wires attached to them
using silver epoxy. Since the samples have submicron thickness and are
thermally anchored to the insulating substrate, there is no evidence of
eddy-current heating in the films during the magnet pulses.

Longitudinal and transverse voltages are measured simultaneously using
two lock-in amplifiers (at $\sim 100$ kHz, using an output time constant =
30 $\mu$sec), whose outputs are recorded on a transient digitizer. The
ohmicity of both longitudinal and transverse voltages were confirmed
down to the lowest temperatures.  Because the contact alignment is
generally good, the transverse voltages are typically an order of
magnitude smaller than the longitudinal voltages.  Nevertheless, at each
temperature, it is necessary to record data from a pair of magnetic-field
pulses of
opposite polarity, where the reported Hall voltage is the asymmetrical
component of the transverse voltage.
(The asymmetrical component of the longitudinal voltages is always
negligible.)

Figure 1(a) shows $\rho_{ab}$ of the underdoped ($x$=0.13) and
overdoped ($x$=0.23) LSCO films.  Several features of the data
reproduce previous measurements on LSCO single
crystals \cite{logT-PRL,x-dep-PRL,MOS}:  The normal state
magnetoresistance is small, evidenced in Fig. 1(a) by the
superposition of the 60 T data (symbols) on
the 0 T data above $T_c$. Also, the normal-state $\rho_{ab}$ of the
underdoped film crosses from metallic behavior above $T_c$ to
strong insulating behavior at low temperatures. In contrast,
$\rho_{ab}$ of the overdoped film stays metallic down to the
lowest experimental temperature \cite{bigrho}. Finally, the
insulating $\rho_{ab}$ of
the underdoped film appears to diverge as $\log (1/T)$ (inset),
the same temperature dependence reported in LSCO single crystals
\cite{logT-PRL,MOS}.

Similar data from three different Bi-2201 crystals is given in Fig.
1(b).  All three crystals have the same nominal composition, the
same $T_c \sim$13 K, and were grown from the same batch in the same
crucible.  The fact that $\rho_{ab}$ differs by more than a factor of
two among these crystals presumably reflects significantly varying
disorder, which is believed to be due to
oxygen vacancies in the ${\rm CuO_2}$ planes \cite{Mackenzie}.
Only sample C, with $\rho_{ab} \leq 100 \mu \Omega$-cm,
remains metallic down to the lowest
temperatures once superconductivity is suppressed with the magnetic
field, a result whith is consistent with a previous report on a
different series of Bi-2201 crystals \cite{Bi2201-PRL}.  The inset to
Fig. 1(b) shows that the weakly insulating behavior in samples
with higher resistivities is consistent with $\log (1/T)$.
This result suggests that the
logarithmic divergence of $\rho_{ab}$ arises, in part, from the
presence of disorder in the samples.

Figure 2 shows Hall data on the same sets of samples.
The inset of Fig. 2(a) shows two traces of the Hall resistivity,
$\rho_{\rm Hall}$, versus magnetic field, $B$, for the underdoped
LSCO film at temperatures above and below $T_c$.
Note that $\rho_{\rm Hall}$ is linear in magnetic field in the
absence of superconductivity, which yields an unambiguous value for
the normal state $R_H$ for these samples. The main panel of Fig. 2(a)
shows the temperature dependence of the normal state $R_H$ for the
two LSCO thin films. Above $T_c$, the underdoped and overdoped
films show broad peaks in $R_H$ at $\sim 80$ K and $\sim 40$ K,
respectively, whose magnitude and positions are
consistent with data in the literature \cite{Hwang}.
The intense magnetic field, however, reveals that both samples exhibit
a nearly temperature-independent $R_H$ at low temperatures (the
data are consistent with a constant $R_H$, although there is
a $\sim 15$\% experimental uncertainty). This common low-temperature
saturation of $R_H$ occurs despite the drastically contrasting
behavior of $\rho_{ab}$ shown in Fig. 1(a). In particular, note
that $\rho_{ab}$ in the underdoped film changes by a factor of two
between 0.8 K and 50 K, while the change in $R_H$ is less than 30\%.

Figure 2(b) shows the normal-state $R_H$ of the three Bi-2201
crystals, along with representative $\rho_{\rm Hall}$ traces in the inset.
Note that, once again, the normal state
$\rho_{\rm Hall}$ depends linearly on $B$,
although the noisier traces from the Bi-2201 single crystals yield larger
error bars in determining $R_H$.
The large temperature dependence of $R_H$ at high temperatures and
the broad peak in $R_H$ near 100 K are, as in the LSCO data, consistent
with previous reports \cite{Mackenzie}.
As with the LSCO thin films, however, $R_H$ of the Bi-2201 crystals
becomes essentially temperature independent at low temperatures, at
least within the experimental resolution.

Although it is not clear how to properly interpret the low-temperature
normal-state $R_H$ of the high-$T_c$ cuprates, it is clear that
the hole-doped LSCO and Bi-2201 behave in stark contrast 
to the electron-doped ${\rm Nd_{2-x}Ce_{x}CuO_{4}}$, in which 
$\rho_{ab}$ is constant yet $R_H$ continues to be strongly temperature
dependent down to low temperatures \cite{NCCO}.
Also, there is no obvious accounting for the data of Figs. 1 and 2
in terms of conventional disordered 2D metals.
As discussed earlier, a logarithmically divergent $\rho_{ab}$
can result from disorder-enhanced electron interactions; however,
this mechanism is effectively ruled out because
the logarithmically divergent $\rho_{ab}$ is not
accompanied by a similarly divergent $R_H$. On the other hand,
while conventional weak localization gives a temperature-independent 
$R_H$, it has
previously been ruled out as a mechanism underlying the observed
$\log (1/T)$ dependence of $\rho_{ab}$ \cite{logT-PRL,Preyer,Jing}.
Further evidence against conventional theories
comes in attempting to interpret the magnitude of the
low-temperature $R_H$ in terms of a carrier density.  Such an approach
gives a carrier number per Cu atom of 0.3 (LSCO $x$=0.13), 0.9
(LSCO $x$=0.23), and 0.6 (Bi-2201).  While
each of these values might suggest a \lq\lq large" Fermi surface in
the cuprates, they are significantly
larger than the carrier concentration expected from the nominal doping.

Among the models specific to high-$T_c$ cuprates which have been
proposed to account for the logarithmic divergence of $\rho_{ab}$ and
$\rho_c$, some include a constant $R_H$ at low temperatures.
In considering disorder effects in 2D Luttinger liquids, Anderson
{\it et al.} \cite{Anderson} discuss a low-temperature regime in which
scattering due to disorder inhibits spin-charge separation.  In this
regime, $\rho_{ab}$ and $\rho_c$ both diverge with the same weak power
law and, more importantly
for this data, the Hall resistivity is expected to be temperature
independent.  In a bipolaron model of high-$T_c$ superconductivity,
a random disorder potential leads to carrier localization marked by
a logarithmically divergent $\rho_{ab}$ and a
constant $R_H$ at low temperatures \cite{Alexandrov}. In the
marginal Fermi liquid theory, the logarithmically divergent $\rho_{ab}$
results from a suppression in the density of states, yet the
$R_H$ will be constant\cite{Varmapc}.

Finally, we discuss our data in terms of the now-well-established
$T^2$ dependence of the cotangent of the Hall angle,
$\cot \Theta_H \equiv \rho_{xx}/\rho_{\rm Hall}$.
This behavior has been
observed in many high-$T_c$ cuprates above $T_c$ and is believed to
be a consequence of two different scattering times for carrier
motion along and perpendicular to the Fermi surface \cite{Hwang,Chien}.
Figure 3 contains plots of $\cot \Theta_H$ at 60 T
versus $T^2$ for our underdoped and overdoped LSCO films \cite{notBi}.
The Hall angle of the overdoped sample does not obey the $T^2$ law,
consistent with Hwang {\it et al.} \cite{Hwang}.  Rather, there is a
slightly sub-$T^2$ behavior which continues down to the lowest
temperatures without any apparent crossover to a low-temperature regime.
In contrast, the underdoped film shows a good
$T^2$ behavior above $\sim$80 K ($\sim$6$\times$10$^3$ K$^2$) and a rapid
increase at lower temperatures.  Note that $\sim$80 K is a
characteristic temperature for this sample:  in Fig. 1(a), it marks
the first deviation of $\rho_{ab}$ from the high-temperature $T$-linear
behavior; while in Fig. 2(a), it marks the maximum in $R_H$.
The break down of the $T^2$ behavior of $\cot \Theta_H$ may
suggest that impurity scattering dominates at low temperatures
and a {\it single} impurity-scattering time dominates the two
separate scattering times which govern the higher temperature behavior.
This is reminiscent of the behavior reported in the heavily overdoped
${\rm Tl_2Ba_2CuO_y}$ system \cite{Mackenzie-Tl}, where lifetime
separation disappears at low temperatures and a single scattering
time with an anomalous temperature dependence governs the carrier
transport.

To summarize,
a 61-T pulsed magnetic field suppresses superconductivity
and reveals an essentially temperature independent normal-state
Hall coefficient, $R_H$, at low temperatures. This behavior is
observed in both underdoped and overdoped LSCO
thin films, despite the fact that $\rho_{ab}$ in these samples
varies from insulating to metallic.  The normal-state $R_H$ is also
found to be nearly temperature-independent in a series of 
Bi-2201 single crystals
in which increasing disorder 
causes the low-temperature $\rho_{ab}$ to cross from metallic to
insulating behavior.

We thank M. Berkowski for growing the ${\rm LaSrAlO_4}$ substrates. We
gratefully acknowledge helpful discussions with E. Abrahams, P.W. Anderson,
A.P. Mackenzie, and C.M. Varma.

%
%

\figure{FIG. 1.  $\rho_{ab}$ in 0 T (solid lines) and 60 T (symbols)
of (a) underdoped ($x$=0.13) and overdoped ($x$=0.23) LSCO thin
films and (b) three La-doped Bi-2201 single crystals with varying
amounts of disorder.  The insets illustrate that the insulating
$\rho_{ab}$ is consistent with a $\log (1/T)$ divergence.
\label{fig1}}

\figure{FIG. 2.  $T$ dependence of the normal state $R_H$ for
(a) the underdoped and overdoped (scaled as indicated) LSCO
thin films and (b) the three La-doped Bi-2201 single crystals.
A typical error bar is shown for each data set.
The insets show examples of the
Hall resistivity traces above and below $T_c$ (The 60 K data in the
inset to (b) is shifted up for clarity).
\label{fig2}}

\figure{FIG. 3.  $\cot \Theta_H$
($\equiv \rho_{xx}/\rho_{\rm Hall}$)
at 60 T versus $T^2$ for the underdoped and overdoped
LSCO films (scaled as indicated).
\label{fig3}}


\begin{references}
\bibitem{bigrho} We note that the $x$=0.23 film is apparently
disordered since $\rho_{ab}$ is roughly a factor of two larger than
single crystals with the same Sr concentration \cite{x-dep-PRL,Kimura}.
This yields a $\rho_{ab}$ in this overdoped film which is comparable to
that of the underdoped film, yet $\rho_{ab}$ remains metallic down to
the lowest temperatures.  This provides further evidence  that the
metallic behavior in the overdoped regime is robust against disorder
and that carrier concentration plays an important role in determining
whether the low temperature behavior is insulating or
metallic \cite{x-dep-PRL}.
\bibitem{notBi}The $\cot \Theta_H$ of Bi-2201 has been known
not to obey the $T^2$ law \cite{Mackenzie}. The complicated $\cot \Theta_H$
temperature dependence exhibited by our Bi-2201 crystals is consistent
with this earlier report.
\end{references}
\end{document}